\documentclass[12pt,letterpaper,english,onecolumn]{IEEEtran}
\usepackage[T1]{fontenc}
\usepackage[latin9]{inputenc}
\usepackage{amsmath}
\usepackage{amssymb}
\usepackage{graphicx}
\usepackage{setspace}
\usepackage[numbers]{natbib}
\makeatletter

\pdfpageheight\paperheight
\pdfpagewidth\paperwidth

\usepackage{algorithmic}
\usepackage[labelsep=period]{caption}

\@ifundefined{showcaptionsetup}{}{%
 \PassOptionsToPackage{caption=false}{subfig}}
\usepackage{subfig}
\makeatother

\usepackage{babel}
\begin{document}

\title{Hypergraph-Based Analysis of Clustered Cooperative Beamforming with
Application to Edge Caching}

\author{Bahar Azari, Osvaldo Simeone, Umberto Spagnolini, and Antonia M.
Tulino}
\maketitle
\begin{abstract}
The evaluation of the performance of clustered cooperative beamforming
in cellular networks generally requires the solution of complex non-convex
optimization problems. In this letter, a framework based on a hypergraph
formalism is proposed that enables the derivation of a performance
characterization of clustered cooperative beamforming in terms of
per-user degrees of freedom (DoF) via the efficient solution of a
coloring problem. An emerging scenario in which clusters of cooperative
base stations (BSs) arise is given by cellular networks with edge
caching. In fact, clusters of BSs that share the same requested files
can jointly beamform the corresponding encoded signals. Based on this
observation, the proposed framework is applied to obtain quantitative
insights into the optimal use of cache and backhaul resources in cellular
systems with edge caching. Numerical examples are provided to illustrate
the merits of the proposed framework.\end{abstract}


\section{Introduction}

A key technology that has been recently introduced in the operation
of wireless cellular systems is cooperative beamforming across non
co-located base stations (BSs). Cooperative beamforming is typically
enabled either by the transmission on backhaul links of common data
streams to a cluster of BSs, or by joint baseband processing carried
out at ``cloud'' processor on behalf of the cluster of BSs \citep{6897914}.
Recently, a new technology has emerged that enables cooperative beamforming
across a cluster of BSs that share the same content by exploiting
the BSs' local storage, namely edge caching \citep{cachemimo}. 

The key idea of edge caching is that of pre-fetching the most requested
files based on their popularity ranking with the goal of decreasing
the number of accesses to the content provider through the backhaul
\citep{6364526}. In this framework, cooperative beamforming is enabled
by storing identical files at nearby BSs \citep{cachemimo,7093176,beam,maddah,DBLP:journals/ftcit/Jafar11}. 

\begin{figure}
\center{\includegraphics[width=11cm]{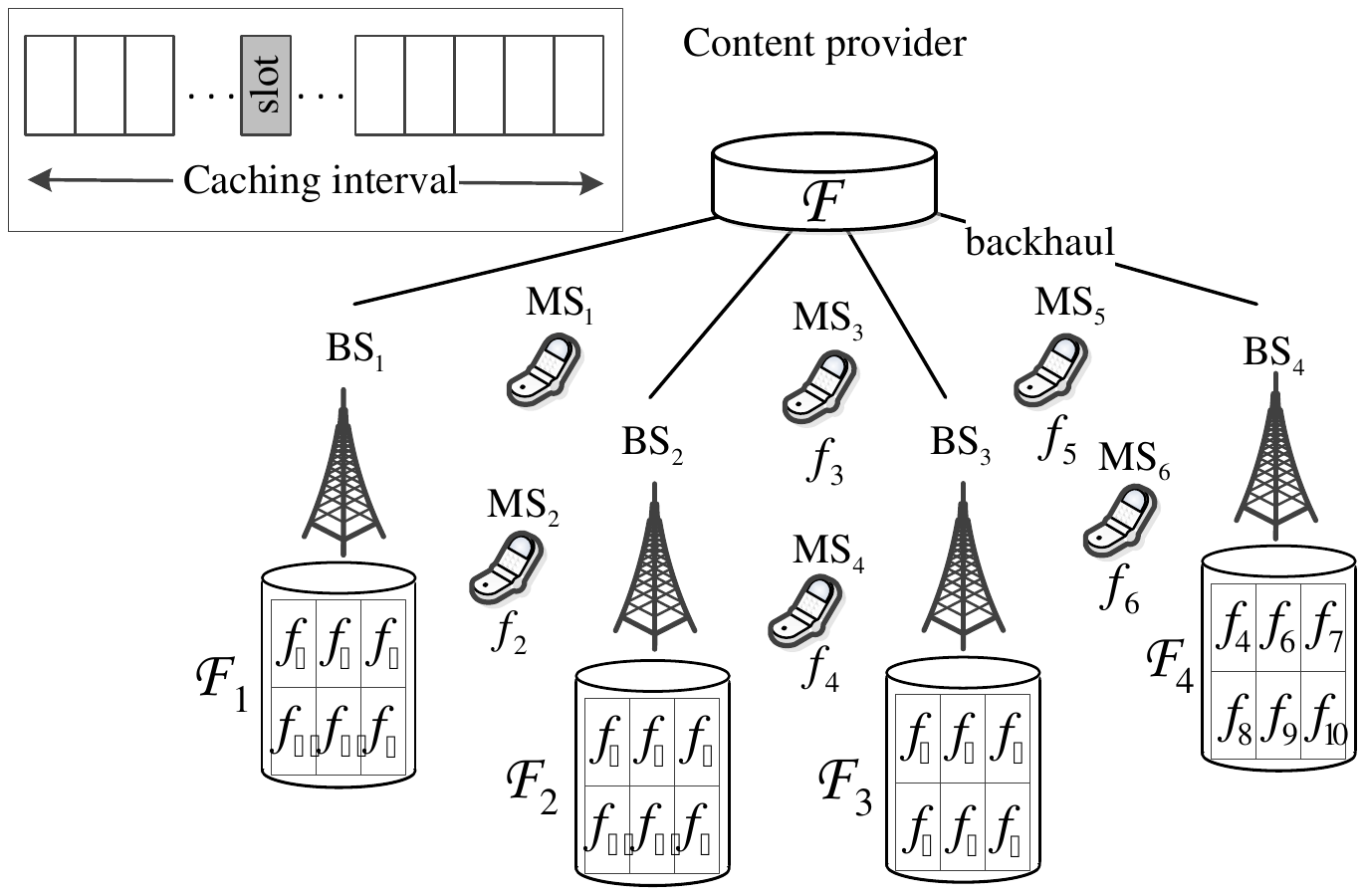}}\protect\caption{Illustration of a system with clustered cooperative beamforming enabled
by edge caching, with indication of the set of files $\mathcal{F}_{m}$
available at each BS $m$.\label{fig:Illustration-of-the}}
\end{figure}

An example of a network that enables clustered cooperative beamforming
is shown in Fig. \ref{fig:Illustration-of-the}. As seen, each content
$f_{k}$ requested by each mobile station (MS) is available at a cluster
of BSs, which can perform cooperative beamforming on the resulting
encoded signal. For instance, content $f_{1}$ is available at the
cluster of BSs $\{\text{BS}_{1},\text{BS}_{2},\text{BS}_{3}\}$ (A
full description of this model in the context of cache-based networks
can be found in Sec. IV). The performance analysis of a system with
a general cluster assignment, including possibly overlapping clusters,
such as in Fig. \ref{fig:Illustration-of-the}, typically requires
solving complex optimization problems (see \citep{6151868} and references
therein).

In contrast, quantitative performance assessment can be carried out
for a number of special message assignments using the high signal-to-noise
ratio (SNR) metric of degrees of freedom (DoF) \citep{DBLP:journals/ftcit/Jafar11}.
Nevertheless, to the best of the authors' knowledge, there is no general
approach to analyze arbitrary message assignments. This letter proposes
a simple framework that aims at accomplishing this goal for a densely
deployed wireless network. The approach is based on a hypergraph formalism
and focuses on the performance of a scheme that uses zero-forcing
beamforming and the orthogonal scheduling of distinct BSs' clusters.

Based on the discussion above, the proposed method can be used to
obtain quantitative insights into the optimal use of backhaul and
caching resources in cache-based wireless networks using the DoF as
performance metric. As a further reference to prior work, we observe
that the joint design of beamforming and backhaul allocation, where
the latter determines which BSs receive each non-cached file on the
backhaul, is studied in \citep{beam} for a fixed pre-defined cache
allocation. In \citep{maddah}, instead, the cache allocation problem
is studied from the point of view of DoF under the assumption that
\textit{all} the requested files are cached at BSs. 

The rest of the letter is organized as follows. Sec. II presents the
system model and Sec. III the proposed hypergraph-based approach.
Sec. IV discuss the application to edge caching and Sec. V provides
some numerical results.

\section{Clustered Beamforming Model}

In this section, we describe the system model for clustered cooperative
beamforming. We consider a wireless network that includes a set ${\cal M}$
of $M$ BSs and a set ${\cal K}$ of $K$ mobile stations (MSs). BSs
and MSs have a single antenna and spatial multiplexing is enabled
only by BSs' cooperation. The extension to case of multiple antennas
is feasible with minor modification but is not covered here. The network
is assumed to be \emph{dense} in the sense that each MS is in the
coverage area of all BSs, i.e., the channel gain from each BS to any
MS is non-zero or, equivalently, the network is fully connected. The
power of each BS is denoted as $P$. Moreover, we assume no time or
frequency diversity so that inteference alignment based on symbol
extentions is not allowed \citep{DBLP:journals/ftcit/Jafar11}.

Each MS $k$ requests a message, or file, $f_{k}$. Each BS $m$ has
available a subset of the requested messages, which we denote as $\mathcal{F}_{m}$
as shown in Fig. \ref{fig:Illustration-of-the}. All the BSs that
have the same message can perform cooperative beamforming for transmission
of the corresponding encoded signal. Note that this assumes the standard
conditions of synchronization and channel state information availability
that are pre-requisites for cooperative beamforming (see, e.g., \citep{6897914}).
We define the set $\mathcal{F}_{\mathrm{r}}=\{f_{k}\in\mathcal{F}:\textrm{ }k=1,..,K\}$
that includes the $K$ messages that are requested by all $K$ MSs. 

Finally, we assume that the MSs are to be served with an equal rate
$R(P)$ {[}bit/s/Hz{]}, hence guaranteeing fairness, where we explicitly
denote the dependence on the transmitted power $P$.

\section{DoF Analysis of Clustered Beamforming\label{sec:Simple} }

In this section, we propose an hypergraph-based framework to evaluate
a high-SNR characterization of an achievable equal rate $R(P)$. We
recall that, if each MS is served at a spectral efficiency $R(P)$
then the corresponding number of DoF per user that are achievable
with the given transmission scheme is defined as \citep{DBLP:journals/ftcit/Jafar11}
\begin{equation}
\text{per-MS DoF}=\underset{P\rightarrow\infty}{\text{lim}}\frac{R(P)}{\text{log}_{2}P}.\label{eq:d}
\end{equation}

\subsection{Cooperative Beamforming Scheme\label{sub:System-Model}}

In order to obtain an achievable DoF metric for arbitrary MSs' requests
and sets $\mathcal{F}_{m},m=1,...,M$, we consider a natural scheme
in which clusters of cooperative BSs, not necessarily disjoint, are
scheduled in orthogonal spectral resources. We adopt such a scheme
for its practicality and simplicity. While an enhanced performance
(\ref{eq:d}) may be generally obtained by means of complex techniques
such as real interference aligment \citep{DBLP:journals/ftcit/Jafar11},
we contend the considered scheme appears to be strongly justified
as any inter-cluster interference would in practice negatively affect
the DoF metric.

To select the cooperative clusters, we note that, if all BSs in a
cluster have all the files requested by an equal number of MSs, then
all the MSs in this subset can be served with no mutual interference
by means of zero-forcing beamforming. This holds under the mentioned
assumption that the network is dense and hence each MS may be served
with non-negligible receiving power by any BS of the set. We define
a subset of MSs as an \textit{independent set} if a subset of BSs
of equal cardinality exists in which all BSs have all the messages
requested by the given set of MSs. MSs in an independent sets can
be served with no mutual interference via zero-forcing beamforming.
We emphasize that, although the clusters of cooperative BSs are not
necessarily disjoint, the independent sets of MSs are non-intersecting.
For example, in Fig. \ref{fig:Illustration-of-the}, the subset \{$\text{MS}_{1},\text{MS}_{2},\text{MS}_{3}$\}
is an independent set because the BSs \{$\text{BS}_{1},\text{BS}_{2},\text{BS}_{3}$\}
all have the files \{$f_{1},f_{2},f_{3}$\} which are requested by
the MSs in this subset.

The scheme at hand then works as follows. In each slot, the MSs are
partitioned into disjoint independent sets, and all independent sets,
along with their corresponding clusters of BSs, are scheduled on orthogonal
time-frequency resources. Therefore, dividing the available time-frequency
resources equally among all the independent sets, a DoF equal to $1/\mathcal{X}$,
where $\mathcal{X}$ is the number of independent sets, can be achieved
on the downlink channel. In the example of Fig. \ref{fig:Illustration-of-the},
beside the independent set \{$\text{MS}_{1},\text{MS}_{2},\text{MS}_{3}$\},
the remaining three MSs cannot be served simultaneously by any subsets
of four BSs. Instead, any two of these MSs can be served by at least
two BSs to form an independent set. Hence, we can take $\left\{ \left\{ \text{MS}_{1},\text{MS}_{2},\text{MS}_{3}\right\} ,\left\{ \text{MS}_{4},\text{MS}_{5}\right\} ,\left\{ \text{MS}_{6}\right\} \right\} $
as subsets defining the desired partition. The resulting per-MS DoF
on the downlink is hence 1/3. 

As a summary, the per-MS DoF achieved by the scheme at hand is given
as 
\begin{equation}
\text{per-MS DoF}=\frac{1}{\mathcal{X}},\label{eq:dof}
\end{equation}
where $\mathcal{X}$ is the number of independent sets of MSs. The
rest of this section is devoted to the calculation of the minimum
number of independent sets $\mathcal{X}$ through a hypergraph coloring
problem.

\subsection{Hypergraph Framework\label{sub:Per-MS-DoF}}

A hypergraph $\mathcal{H}=(\mathcal{K},\mathcal{E})$, where the vertex
set $\mathcal{K}$ is the set of MSs and $\mathcal{E}$ is the set
of hyperedges associated to a given allocation of files across the
BSs and to a set of MSs' requests. A hyperedge $e\subseteq\mathcal{K}$
is in $\mathcal{E}$ if there is no subset of $\left|e\right|$ BSs
such that each BS in the subset has all the files requested by the
MSs in the set $e$. Under this definition, the independent sets introduced
above correspond exactly to the independent sets of $\mathcal{H}$.
We recall in fact that an independent set of a hypergraph $\mathcal{H}$
is a subset of the vertex set such that no subset of this set is a
hyperedge of $\mathcal{H}$ \citep{bookhyper}. 

\begin{figure}
\center{\subfloat[\label{fig:2a}]{\includegraphics[width=6cm]{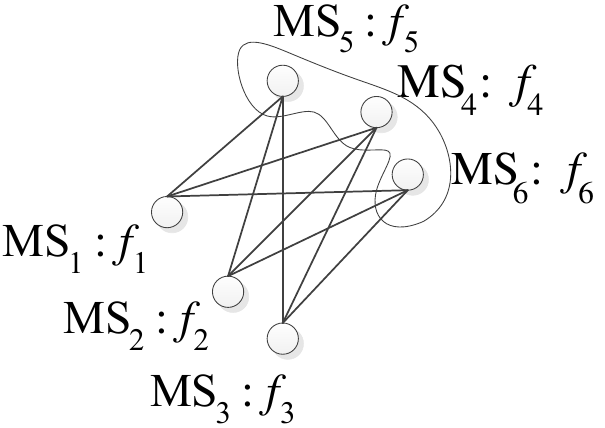}

} \subfloat[\label{fig:2b-1}]{\includegraphics[width=6cm]{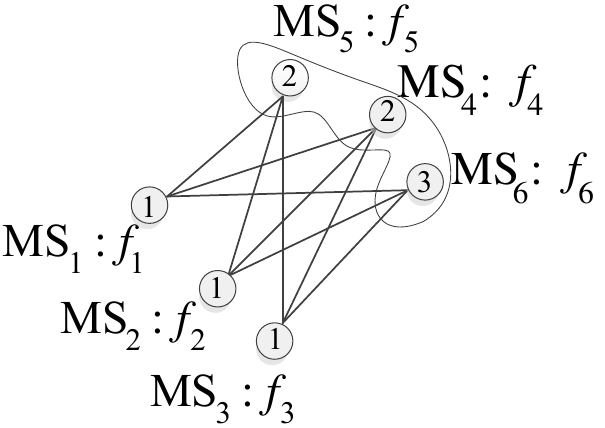}

}}

\protect\caption{(a) Illustration of hypergraph $\mathcal{H}$ associated with the
configuration in Fig. \ref{fig:Illustration-of-the}. Hyperedges of
cardinality 2 are represented as regular edges, while the hyperedges
of larger cardinality are represented as closed curves including a
set of vertices. (b) Illustration of a colored hypergraph, solved
with a standard greedy coloring, associated to the MSs in Fig. \ref{fig:Illustration-of-the},
where numbers on the vertices indicate colors.}
\end{figure}

We focus with no loss of generality of simple hypergraphs in which
only minimal hyperedges are included in $\mathcal{E}$ \citep{bookhyper}.
A hyperedge $e\in\mathcal{E}$ is minimal if no subsets of $e'\subseteq e$
is an hyperedge. In our context, this implies that, if a hyperedge
of cardinality $\left|e\right|$ exists, then all subsets of $e$
with cardinality less than $\left|e\right|$ correspond to independent
sets and hence zero-forcing joint beamforming is possible within any
such subset. In Fig. \ref{fig:2a} shows the hypergraph associated
to the network configuration in Fig. \ref{fig:Illustration-of-the}
considering only the files in the BSs' caches. 

The minimum number of independent sets of $\mathcal{H}$ is known
as the chromatic number $\mathcal{X}$ and corresponds exactly to
the number of independent sets of MSs introduced above. Calculating
the chromatic number $\mathcal{X}$ requires solving the hypergraph
coloring problem, which is known to be NP-hard \citep{bookhyper}.
The hypergraph coloring problem consists of the assignment to each
vertex of a hypergraph $\mathcal{H}$ a color of such no hyperedge
is monochromatic. Here we resort to a standard greedy coloring algorithm
as detailed in, e.g., \citep{bookhyper}. The hypergraph coloring
problem consists of the assignment of a color to each vertex of a
hypergraph $\mathcal{H}$, such that no hyperedge is monochromatic.
Note that colors are identified by integer numbers and that $\mathcal{X}$
equals to the maximum number used to color the vertices of hypergraph.
Once coloring is completed, subsets of vertices with the same color
form disjoint independent sets and hence should be scheduled in different
spectral resources. An example is shown in Fig. \ref{fig:2b-1} in
which we fix the permutation $i_{1},i_{2},\ldots,i_{\mathcal{K}}$
as $\text{MS}_{1},\text{MS}_{2},\text{MS}_{3},\text{MS}_{4},\text{MS}_{5},\text{MS}_{6}$
and apply the greedy coloring algorithm in \citep{bookhyper}.

\section{Application to Edge Caching}

The framework described above can be applied to obtain quantitative
insights into optimal use of cache and backhaul resources. This application
is discussed in this section.

\subsection{Edge Caching Model}

In a cache-based wireless network, each BS $m$ is endowed with a
cache that can store a set of $N$ files and is connected to the content
provider via a backhaul link that can deliver up to $\mathcal{C}\text{log}P$
bit/s/Hz, where $\mathcal{C}$ is the backhaul capacity measured in
terms of DoF. The files are selected from a set $\mathcal{F}$ of
popular files that remains constant for a period referred to as \textit{caching
interval}. We label the files in decreasing order of popularity in
a given caching interval, so that the file set is indicated as $\mathcal{F}=\{1,2,\ldots,F\}$,
where $f\in F$ is the $f$th most popular file. We define as $\mathcal{F}_{\text{c},m}$
the set of files cached by BS $m$ for a given caching interval.

Time is characterized by two different scales so that each \textit{caching
interval} contains \textit{multiple slots}, as shown in the inset
of Fig. \ref{fig:Illustration-of-the}. At any slot\emph{, }each MS
$k$ requests a file $f_{k}\in\mathcal{F}$ independently of the others
according to the classical Zipf popularity distribution with parameter
$\gamma\geq0$, i.e., with probability $P_{f}=f^{-\gamma}\diagup\sum_{f=1}^{F}f^{-\gamma}$.
For any given slot, we define the set $\mathcal{F}_{\mathrm{r}}=\{f_{k}\in\mathcal{F}:\textrm{ }k=1,..,K\}$
that includes the $K$ files that are requested by all MSs. Note that
any file in the set $\mathcal{F}_{\mathrm{r}}\backslash\mathcal{F}_{\mathrm{c}}=\cup_{m=1}^{M}\mathcal{F}_{\text{c},m}$,
if requested, needs to be downloaded on the backhaul links of some
BSs in order to be transmitted to the requesting MSs. We also observe
that the equal rate $R(P)$ at which transmission is possible in a
slot, generally changes from slot to slot due to the varying MSs\textquoteright{}
requests.

\subsection{DoF Analysis of Caching and Backhaul Policies}

The design of the edge caching system requires the definition of the
policies used to populate the caches and to allocate the backhaul
resources. The \textbf{caching allocation policy}, which is to be
applied at the long time-scale of caching intervals, determines the
subset $\mathcal{F}_{\text{c},m}$ of files that each BS $m$ in the
network caches for a caching interval. Instead, the \textbf{\textit{\emph{backhaul
allocation policy }}}\textit{\emph{is applied in each slot}} to determine
which files in the set $\mathcal{F}_{\mathrm{r}}\backslash\mathcal{F}_{\mathrm{c}}$
of files that are requested but not in the caches should be sent on
each backhaul link to the connected BS. We next provide some examples
of basic caching and backhaul policies that will be considered in
the numerical results of Sec. V.

\subsubsection*{Examples of caching policies}

(\emph{i}) Cache Most Popular (CMP): All the BSs pre-fetch the $N$
most popular files for caching. Note that CMP enables full cooperation
among the BSs on the transmission of the cached files but may cause
a significant number of cache misses if $N$ is small. (\emph{ii})
Cache Distinct (CD): For a given arbitrary order of the BSs, the first
BS stores the $N$ most popular files, the second BS the next $N$
most popular files, and so on, in such a way that, assuming that $MN\leq F$,
there is no duplication of files in the BSs' caches. CD makes joint
beamforming impossible, at least based solely on the cached files,
but it minimizes the probability of a cache miss. (\textit{iii}) Hybrid
Caching (HC): All BSs cache the same $N_{\text{CMP}}\leq N$ most
popular files to induce cooperative transmission, and then different
BSs cache distinct files from the rest of the files following the
CD policy. Note that HC generalizes both CMP and CD.

\subsubsection*{Example of backhaul policy}

Due to space constraint, we mention here on the simple Greedy Download
(GD) policy. This policy allocates each missing file in $\mathcal{F}_{\mathrm{r}}\backslash\mathcal{F}_{\mathrm{c}}$,
following some pre-specified order (here, the order of popularity),
to a subset of BSs that yields the largest per-MS DoF detailed below.

\begin{figure}
\center{\includegraphics[width=7.8cm]{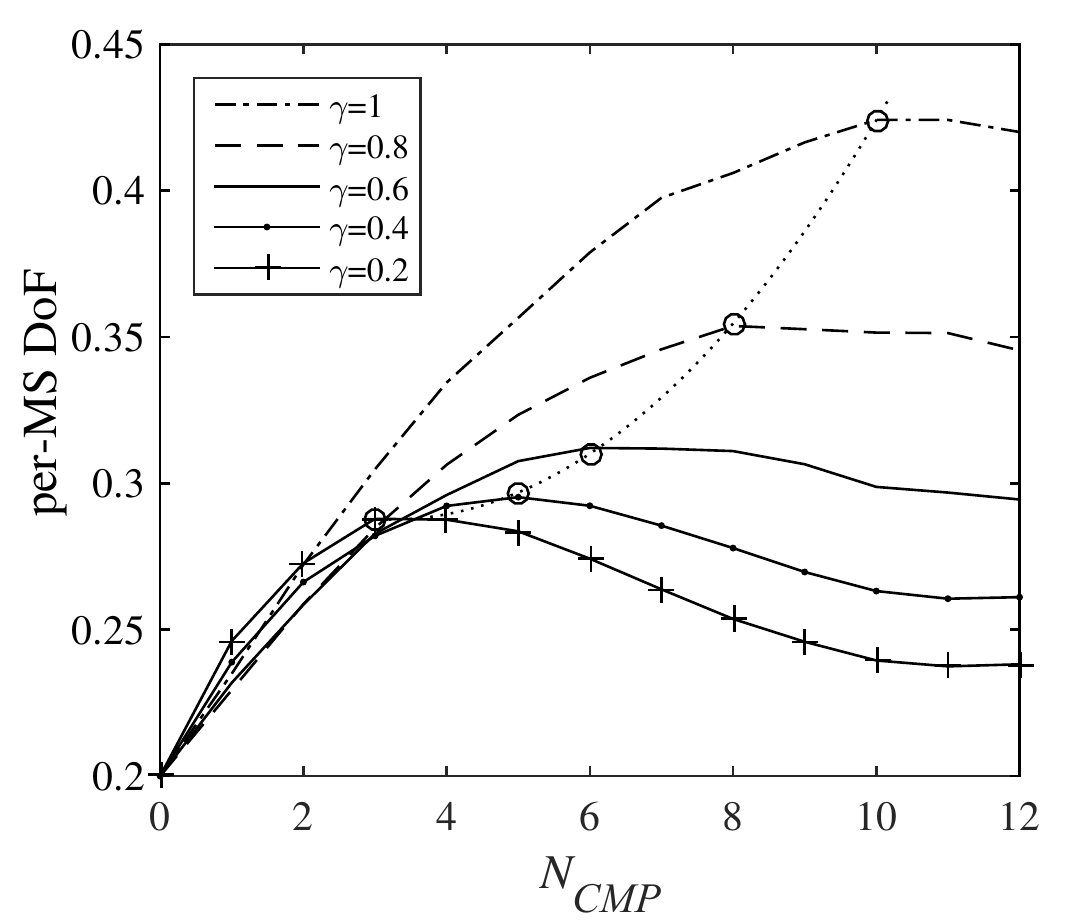}}

\protect\caption{Per-MS DoF of the HC caching policy under GD backhaul policy for different
popularity distribution exponents $\gamma$ ($N=12,M=K=5,F=60$).\label{fig:hybrid}}
\end{figure}

For any given caching and backhaul policy, at any slot, the model
reduces to the one studied in the previous sections in which each
MS requests a file and each BS $m$ has available a subset $\mathcal{F}_{m}$
of the files. Note that the set $\mathcal{F}_{m}$ contains both the
files in $\mathcal{F}_{\text{c},m}$ that are present in the cache
of BS $m$ and also the files that have been downloaded on the corresponding
backhaul link. Therefore, we can adopt the framework developed in
Sec. III in order to obtain an assessment of the per-MS DoF in any
given slot. To this end, denote as $F_{max}$ the maximum number of
files transmitted on a backhaul link to any BS by the given backhaul
transmission policy. The per-DoF is then limited, not only by the
downlink DoF studied in Sec. III, but also by $\mathcal{C}/F_{max}$.
This is because the equal rate $R(P)$ cannot be larger than the rate
supported on the backhaul link if at least one of the files needs
to be downloaded from the content provider. The per-MS DoF achieved
in a given slot is then evaluated as 
\begin{equation}
\text{per-MS DoF}=\text{min}\left(\frac{\mathcal{C}}{F_{max}},\frac{1}{\mathcal{X}}\right),\label{eq:DoF and Backhaul}
\end{equation}
where $\mathcal{X}$ is the chromatic number of the hypergraph corresponding
to the configuration of MSs' requests and caches in the given slot.
The performance of specified caching and backhaul policies can then
be assessed by averaging (\ref{eq:DoF and Backhaul}) over the randomness
of the MSs' requests.

\section{Numerical Results\label{sub:Caching-Policy}}

In this section, we provide some numerical example in order to illustrate
the type of conclusions that can be obtained by means of the proposed
framework. We emphasize again that the results can be obtained in
a straightforward manner by implementing the hypergraph coloring approach
discussed in Sec. III. We focus on edge caching and show in Fig. \ref{fig:hybrid}
the average per-MS DoF, obtained from (\ref{eq:DoF and Backhaul}),
for the HC caching policy and GD backhaul policy versus the number
of files $N_{\text{CMP}}$ to be stored at all caches. Different curves
are obtained by varying the popularity exponent $\gamma$. For a larger
$\gamma$, allocating a bigger part of the cache to the same files,
as measured by $N_{\text{CMP}}$, yields a higher per-MS DoF, as this
maximize the cooperation opportunities without causing too many cache
misses. In particular, Fig. \ref{fig:hybrid} enables the quantitative
estimate of the optimal value of $N_{\text{CMP}}$, as indicated by
the dotted line. 

\begin{figure}
\center{\includegraphics[height=7cm]{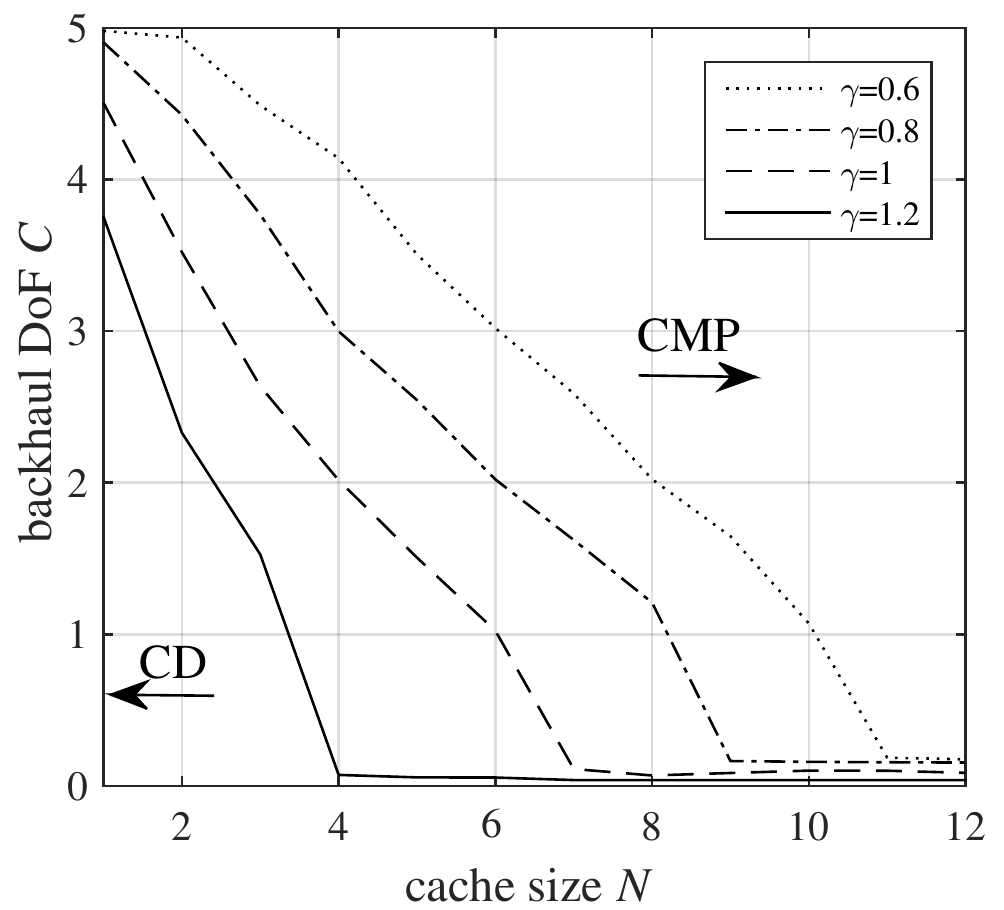}}

\protect\caption{Regions of backhaul DoF $\mathcal{C}$ and cache size $N$ values
in which CMP (right of the curve) or CD (left of the curve) outperforms
the other under GD backhaul policy for different popularity distribution
exponent $\gamma$ ($M=K=5,F=60$).\label{fig:locus}}
\end{figure}

To get more insight into the comparison between the CMP ($N_{\text{CMP}}=N$)
and CD ($N_{\text{CMP}}=0$) caching policies, Fig. \ref{fig:locus}
represents the regions of backhaul DoF $\mathcal{C}$ and cache size
$N$ values in which the CMP or CD caching policy outperforms the
other under the GD backhaul policy as a function of the popularity
exponent $\gamma$. The figure allows to obtain the values of $N$
above which CMP is advantageous for a fixed $\mathcal{C}$, or, for
a fixed $N$, the values of $\mathcal{C}$ that are sufficiently large
to compensate for the cache miss events. 

\bibliographystyle{ieeetr}
\bibliography{REFNEW}

\begin{thebibliography}{1}

\bibitem{6897914}
A.~Checko and et~al., ``Cloud {RAN} for mobile networks{--A} technology
  overview,'' {\em IEEE Communications Surveys Tutorials}, vol.~17, no. 1,
  pp.~405--426, First quarter 2015.

\bibitem{cachemimo}
A.~Liu and V.~Lau, ``Exploiting base station caching in {MIMO} cellular
  networks: Opportunistic cooperation for video streaming,'' {\em IEEE Trans.
  Signal Process.}, vol.~63, no. 1, pp.~57--69, Jan. 2015.

\bibitem{6364526}
N.~Golrezaei, K.~Shanmugam, A.~Dimakis, A.~Molisch, and G.~Caire, ``Wireless
  video content delivery through coded distributed caching,'' in {\em Proc.
  IEEE ICC 2012}, pp.~2467--2472, Ottawa, ON, Jun. 2012.

\bibitem{7093176}
W.~Han, A.~Liu, and V.~Lau, ``Degrees of freedom in cached mimo relay
  networks,'' {\em IEEE Trans. Signal Process.}, vol.~63, no.15,
  pp.~3986--3997, Aug. 2015.

\bibitem{beam}
X.~Peng, J.~C. Shen, J.~Zhang, and K.~B. Letaief, ``Joint data assignment and
  beamforming for backhaul limited caching networks,'' in {\em Proc. IEEE PIMRC
  2014}, Washington, DC, Sep. 2014.

\bibitem{maddah}
M.~A. Maddah-Ali and U.~Niesen, ``Cache-aided interference channels,'' in {\em
  Proc. IEEE ISIT 2015}, Hong Kong, China, Jun. 2015.

\bibitem{DBLP:journals/ftcit/Jafar11}
S.~A. Jafar, ``Interference alignment: {A} new look at signal dimensions in a
  communication network,'' {\em Found. Trends Commun. Inf. Theory}, vol.~7,
  no.1, no.~1, pp.~1--136, 2011.

\bibitem{6151868}
S.~Kaviani, O.~Simeone, W.~Krzymien, and S.~Shamai, ``Linear precoding and
  equalization for network mimo with partial cooperation,'' {\em IEEE Trans.
  Veh. Tech.}, vol.~61, no. 5, pp.~2083--2096, Jun 2012.

\bibitem{bookhyper}
A.~Bretto, {\em {Hypergraph theory. An introduction.}}
\newblock Mathematical Engineering. Cham: Springer, 2013.

\end{thebibliography}

\end{document}